\documentclass{article}
\usepackage{arxiv}

\usepackage[utf8]{inputenc} % allow utf-8 input
\usepackage[T1]{fontenc}    % use 8-bit T1 fonts
\usepackage{hyperref}       % hyperlinks
\usepackage{url}            % simple URL typesetting
\usepackage{booktabs}       % professional-quality tables
\usepackage{amsfonts}       % blackboard math symbols
\usepackage{nicefrac}       % compact symbols for 1/2, etc.
\usepackage{microtype}      % microtypography
\usepackage{lipsum}		% Can be removed after putting your text content
\usepackage{graphicx}
\usepackage{natbib}
\usepackage{doi}
\usepackage{graphicx}

\title{Research on Virus Cyberattack-Defense Based on Electromagnetic Radiation}

\date{} 					% Or removing it

\author{ \href{https://orcid.org/0000-0003-0852-424X}{\includegraphics[scale=0.06]{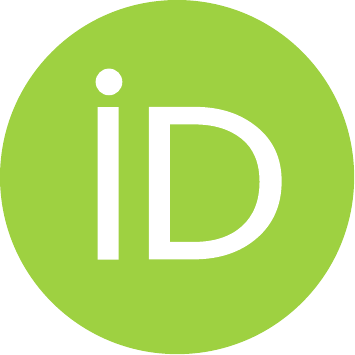}\hspace{1mm}Ruochen Wu}\thanks{Ruochen Wu is currently a Ph.D. candidate with research interests in signal processing, radar remote sensing, and cyber attack-defense countermeasure.} \\
	Dept. of Signal Theory and Communications\\
	Universitat Politècnica de Catalunya\\
	C/ Jordi Girona 1-3, Barcelona 08034, Spain \\
	\texttt{ruochen.wu@upc.edu} \\
}

\renewcommand{\headeright}{}
\renewcommand{\undertitle}{}
\renewcommand{\shorttitle}{Research on Virus Cyberattack-Defense Based on Electromagnetic Radiation}

\hypersetup{
pdftitle={Research on Virus Cyberattack-Defense Based on Electromagnetic Radiation},
pdfsubject={SP},
pdfauthor={Ruochen Wu},
pdfkeywords={Cyberattack, Cyberweapon, Signal processing, Radio signal, Radiation injection},
}

\begin{document}
\maketitle

\begin{abstract}
Information technology and telecommunications have rapidly permeated various domains, resulting in a significant influx of data traversing the networks between computers. Consequently, research of cyberattacks in computer systems has become crucial for many organizations. Accordingly, recent cybersecurity incidents have underscored the rapidly evolving nature of future threats and attack methods, particularly those involving computer viruses wireless injection. This paper aims to study and demonstrate the feasibility of remote computer virus radiation injection. To achieve this objective, digital signal processing (DSP) plays a vital role. By studying the principles and models of radiation attacks and computer virus propagation, the modulation of the binary data stream of the simulated virus into a terahertz radar carrier signal by Phase-Shift Keying (PSK) is simulated, enabling the implementation of an attack through the "field to line" coupling of electromagnetic signals. Finally, the defense and countermeasures based on signal recognition are discussed for such attacks. Additionally, an idea of establishing a virus library for cyberattack signals and employing artificial intelligence (AI) algorithms for automated intrusion detection is proposed as a means to achieve cybersecurity situation awareness.
\end{abstract}

% keywords can be removed

\section{Introduction}
In the information age, cybersecurity is an increasingly serious issue. With the advancement of computer and communication technology, network attacks represented by computer viruses have become a serious threat to the majority of users. Since the 1980s, there have been computer viruses that pose a serious threat to computer security \citep{cohen1987computer}. Currently, network devices, such as computers, are extensively utilized in various fields. As a result, information security and network attacks have become modern methods of warfare. Especially in high-tech warfare environments, cyber attack-defense combat capabilities will be the decisive factor in determining the outcome of the information battlefield \citep{hu2010research,kostyuk2019invisible}. With the substantial increase in network transmission distance and speed, the effectiveness of electromagnetic attacks has also increased accordingly. Compared to wired attack and defense technologies, wireless attacks, such as electromagnetic radiation, are more covert and challenging to prevent.

The attack methods of computer viruses mainly include wireless injection \citep{rieback2006your}, wired insertion \citep{schmitt2002wired}, network intrusion \citep{kumar2007survey}, mail transmission \citep{newman2002email}, and node attack \citep{upadhyay2020modeling}. Among these methods, wireless injection is an important means in space information countermeasure. Additionally, in the field of information countermeasure, research on computer virus weapons is currently a trending topic. The goal is to radiate and attack the vulnerable areas of the enemy's information system with electromagnetic radiation information containing computer viruses, thereby injecting viruses that can paralyze the information systems \citep{zhangkai2018}. Wireless injection involves converting computer viruses into a virus code data stream, that is, radio signal, and then modulate it into electromagnetic waves and transmitted via radio signals. This allows the virus code to be radiated into the enemy radio receiver through transmitter, exploiting any loopholes or weak links in their system to gain entry. In the future, computer virus intrusion methods will focus on injecting viruses through antenna electromagnetic waves transmitted by antennas and using satellite radiation injection. As early as the early 21st century, the U.S. Department of Defense began developing a virus cannon capable of injecting computer viruses over long distances \citep{li2005}. The virus is radiated to the enemy's host computer, sensors, and network bridges through electromagnetic waves, thereby waiting for an opportunity to attack and destroy the enemy's weapon system, command and control system, and communication system.

This paper presents a comprehensive study on the computer virus radiation attack-defense technology, which systematically explains the computer virus radiation attack mechanism, radiation injection principle, and signal modulation and processing technology. The theoretical model demonstrates the feasibility of implementing virus electromagnetic radiation strikes, which lays the foundation for information security and cyber attack-defense countermeasure and related signal processing. In addition, the fundamental framework of binary data stream modulation to the electromagnetic signal is studied, and the operational mode that enables the coupling of the radiation signal to execute an attack is determined.

\section{Theoretical Background and Methods}
Computer virus attack-defense is an important research topic in the field of cybersecurity. For the radiation injection method, virus propagation model, radiation injection framework, and signal modulation principle are the foundation for information countermeasure missions.

\subsection{Mathematical model of computer virus propagation}
To initiate a computer virus infection, the virus must be injected into a specific computer within the local area network (LAN). Once injected, the virus can spread and infect multiple devices during the process of information communication, allowing to carry out the attack. Researchers have established relevant mathematical models in order to study computer virus network attacks and their infection patterns \citep{guo1999}.

Suppose there is a computer group consisting of $N$ computers that are threatened by virus $V$, and there is data interaction within the network. $N$ represents the maximum number of computers that the virus can infect. The set $C=\left \{C_1, C_2, ..., C_N\right \}$ represents these $N$ computers. Define the binary relationship $D_{C}=\left \{\left \langle C_{i},C_{j}\right \rangle \mid C_{i},C_{j}\in C,i\ne j\right \}$ on $C$ according to the data communication relationship between them, where there is data flow from $C_{i}$ to $C_{j}$; $C_{i}$ is the source computer device and $C_{j}$ is the target device. Define $M_{V}=\left \{C_{i}\mid C_{i}\in C\right \}$ as the set of computers infected with viruses, where $C_{i}$ represents a computer infected with virus. It is stipulated that from the moment the virus enters the system ($n=0$), the number of the infected computer is $C_{1}$. The condition for a computer $C_{j}$ ($j\ne 1$) to be infected with the virus is: $\left ( \left \langle C_{i},C_{j}\right \rangle \in D_{C}\right )\cap \left ( C_{i}\in M_{V}\right )\cap \left ( C_{i}\notin M_{V}\right )$.

Let $X_{n}$ represent the number of computers infected with viruses in $C$ at the nth unit time, therefor, $\left \{X_{n},n\ge 1\right \}$ constitutes a discrete random process. At the $n$th unit time, the number of virus-infected computers is $E\left (X_{n}\right )$, then the number of uninfected computers is $N-E\left (X_{n}\right )$. In the $M$ times of data communication during the $n$th and $n+1$th time intervals, the mathematical expectation of infected computers in $C_{i}$ is $\frac{M}{N}E\left (X_{n}\right )$, while the number of virus-free computers in $C_{j}$, $M-\frac{M}{N}E\left (X_{n}\right )$. According to the randomness of virus propagation, the mathematical expectation of newly infected computers after this copy is:
\begin{equation}
E\left (X_{n+1} \right )-E\left (X_{n}\right )=\frac{M}{N}E\left (X_{n}\right )\left (1-\frac{E\left (X_{n}\right )}{N}\right )
\label{eq1}
\end{equation}
Let $E\left (X_{n}\right )$ be regarded as the discretized value of a continuous function $f\left (x\right )$ at the point $x=n$. According to Eq. \ref{eq1}, the differential equation of $f\left (x\right )$ can be obtained:
\begin{equation}
\frac{df\left (x\right )}{dx}=\frac{M}{N}f\left (x\right )\left (\frac{1-f\left (x\right )}{N}\right )
\label{eq2}
\end{equation}
Because $f\left (x\right )=1$, the number of computers infected with viruses at time $n$ can be obtained by solving and discretizing the separated variables in Eq. \ref{eq2}.
\begin{equation}
E\left (X_{n}\right )=\frac{N}{1+\left (\frac{N}{X_{0}}-1\right )e^{-n\frac{M}{N}}}=\frac{N}{1+\left (N-1\right )e^{-n\frac{M}{N}}}
\label{eq3}
\end{equation}

\begin{figure}[h]
\centering
\includegraphics[scale=0.6]{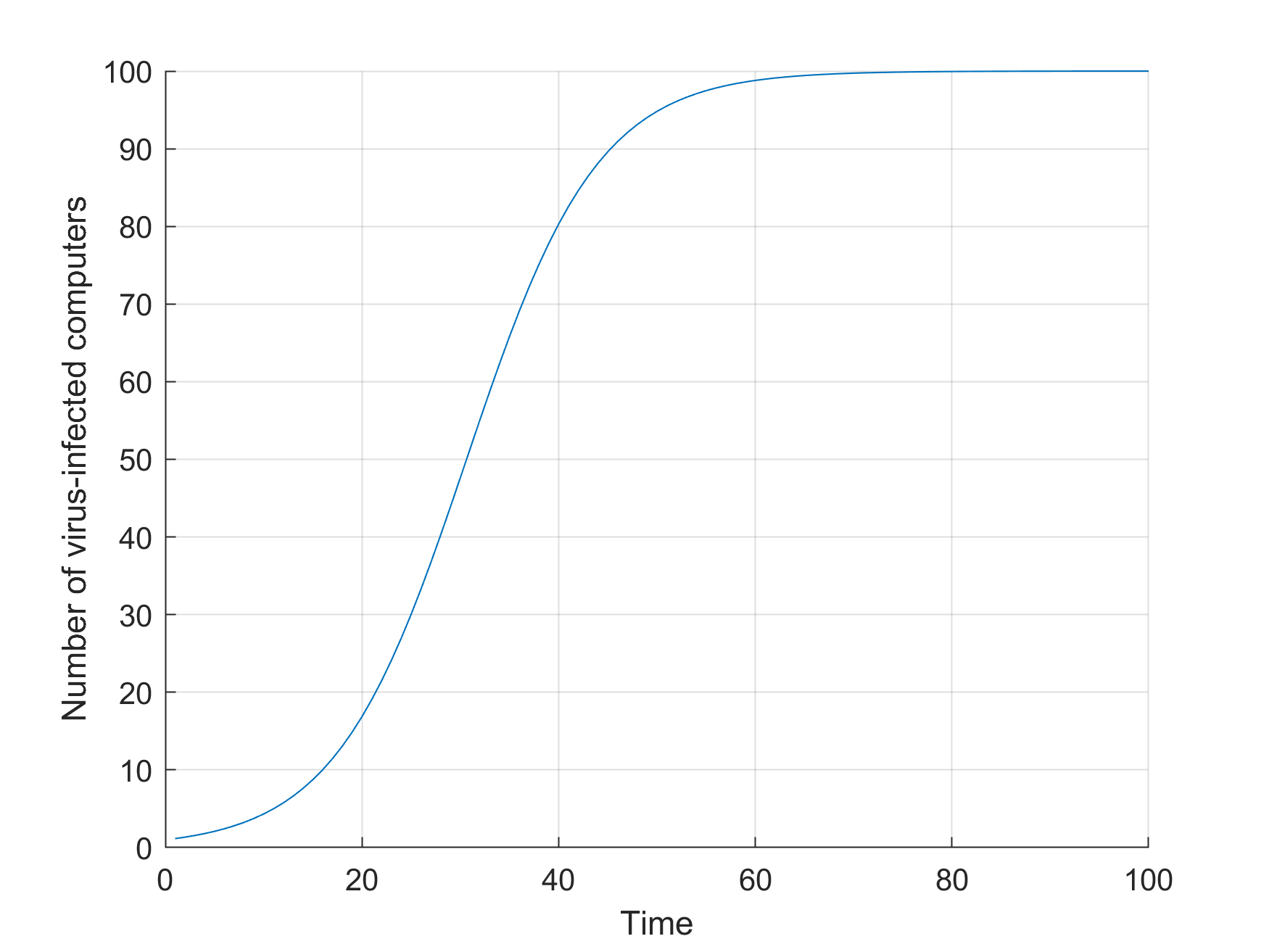}
\caption{Computer virus propagation simulation curve}
\label{fig1}
\end{figure}

Assuming that there are 100 computers connected to each other, the number of data communications between them is 15 times, and one of these computers is infected with a virus. Figure \ref{fig1} shows the computer virus propagation simulation curve based on Eq. \ref{eq3}. Around the 20th unit time, the virus spreads slowly, which is the incubation period. From approximately the 20th to the 60th, the virus shows a rapid growth trend. At this time, the virus is spreading extensively through computer networks due to the interconnected communication in the LAN. After the 60th, the trend of virus spread tends to stabilize because the increasing number of infected computers leads to a rise in network communication traffic, and the virus significantly impacts the performance of the computer itself.

\subsection{Computer virus radiation injection technology}
\subsubsection{Principle}
The process of computer virus radiation injection involves injecting virus information into the network cable through coupling. Once recognized, received, and executed by the computer system, the virus utilizes its rapid and widespread characteristics to launch attacks and disrupt the computer network. In contrast, a wide area network (WAN) is primarily connected through fiber optic cables, which offer robust electromagnetic information security and anti-interference capabilities \citep{jie2007}. The connection mode adopted by LAN makes its radiation coupling ability more strong, making it the primary target of radiation attacks.

Taking Ethernet that implements the IEEE802.3 protocol as an example \citep{archimbaud1992ethernet}, if it is required to successfully inject virus information into it, it should be injected when the network communication is idle. If there are already ongoing or pending conversations in the network, it is necessary to use high-power radiation to force injection \citep{cui2021}. Furthermore, the electromagnetic signal progressively weakens as the communication distance increases. Therefore, in order to ensure successful injection of virus information, the equipment emitting radiation signals must have sufficient power. This makes it possible to use high-power sensors such as satellites and radars as sources of emissions.

\subsubsection{Method}
Based on the fundamental principle of radiation injection, one implementation method involves using a high-power computer virus microwave launcher or a corresponding device to precisely control the peak value of its electromagnetic pulse. This allows for the injection of a virus into a specific part of the enemy's computer system, thereby infecting it. Additionally, the dual modulation technology of high-power microwave and computer virus can be directly combined to transmit a continuous stream of high-power microwave, which is modulated with a computer virus. This approach enables the virus to be injected into a computer that is currently receiving information.

\begin{figure}[h]
\centering
\includegraphics[scale=0.55]{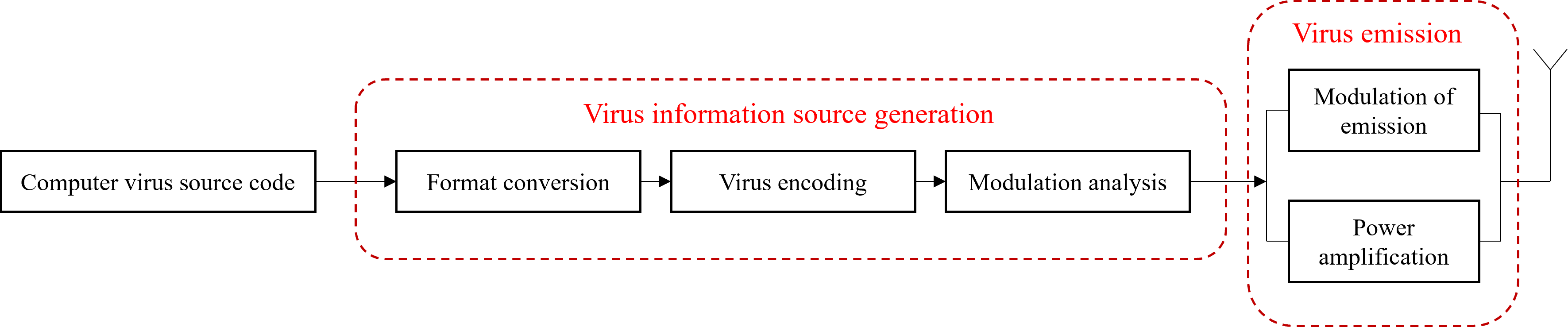}
\caption{Computer virus radiation injection schema}
\label{fig2}
\end{figure}

Figure \ref{fig2} shows the computer virus radiation injection schema, including the process of computer virus signal processing. By analyzing the data format, encoding, and modulation mode of the target network, the virus source code is transformed to match the same format as the enemy network data information. Finally, signal modulation is performed on the coded and modulated virus information source, which is emitted as an electromagnetic wave through the transmitting antenna to attack the enemy's computer system with radiation. Among them:

\begin{itemize}
\item Modulation of emission: signal modulation is performed on the generated virus source and loaded into the original emission signal.
\item Power amplification: Complete output power generation and preamplification.
\end{itemize}

This paper focuses on the relevant methods and technologies of virus signal processing to explore the feasibility of modulating virus information into the emitted signal.

\subsection{Signal modulation technology}
Signal modulation is the process of modifying one or more characteristics of a periodic waveform, known as the carrier signal, by incorporating a modulated signal that typically carries the information to be transmitted \citep{chan1989identification}. In layman's terms, modulation is to move the signal (original information) to be transmitted to the carrier signal. Signal modulation is divided into two categories: analog modulation and digital modulation. In addition, depending on the modulated objects, modulation can be classified into four types: frequency modulation (FM), amplitude modulation (AM), phase modulation (PM), and quadrature amplitude modulation (QAM) \citep{azzouz2013automatic}.

To modulate a computer virus signal, which is converted into a binary data stream, for radar emission, digital modulation is the preferred choice. Digital modulation involves converting discrete digital signals into continuous analog signals through modulation techniques \citep{xiong2006digital}. Various schemes can be chosen based on specific requirements, such as amplitude shift keying(ASK), frequency shift keying (FSK), and phase shift keying (PSK). It allows the transformation of digital signal information into specific characteristics of an analog signal, such as frequency, phase, or amplitude, enabling transmission in a communication system. In ASK, the binary digital signal utilizes $0$ and $1$ to determine the presence or absence of the carrier amplitude, resulting in changes in the carrier amplitude along with the signal. When the signal is $1$, the carrier signal is transmitted, while it is not transmitted when the signal is $0$. FSK allows for the modulation of the carrier frequency based on the digital signal. It involves transmitting carrier signals of different frequencies for binary $1$ and $0$. In this case, $s_{0}\left (t\right )=A\cos \left (2\pi f_{0}t\right )$ is used to represent $0$, and $s_{1}\left (t\right )=A\cos \left (2\pi f_{1}t\right )$, to represent $1$. PSK represents digital binary data using the phase of an analog carrier wave, which varies according to the binary input. When the input signal is $1$, the output is a carrier wave with a phase of $0$. Conversely, when the input signal is $0$, the output is a carrier wave with a phase of $\pi$.

Based on the aforementioned theories and technologies, it is important to select different modulation techniques based on specific application requirements and system design. Computer virus is transmitted in wireless channels as signals encoded in the form of electromagnetic waves. In this case, the receiving end can become infected with the virus information through "field to line" coupling, thereby executing an attack on the target computer. This technique enables wireless signal transmission from a transmitting device to a target device through electromagnetic radiation without physical connection. Within a certain range, it allows signal transmission and injection without direct physical contact with the target device. Transmitting antennas emit radio signals using electromagnetic radiation, creating an electromagnetic field that induces signal information into the target device, enabling signal injection.

\subsection{Signal recognition for computer virus defense}\label{sig recogn}
With the advancement of cyberspace countermeasure, the modulation and encoding techniques of computer virus injection signals are constantly evolving, and the variety of transmitting antennas is becoming increasingly intricate. In order to accurately detect and defend against virus intrusions, the utilization of signal recognition technology is of paramount importance. The underlying principle involves conducting time-frequency analysis on the received radiation signal, extracting its time-domain and frequency-domain features through Fourier transform, and employing a recognition classifier for identification. Once the radiation spectrum of the signal exhibits a significant correlation with the established criteria, it confirms the presence of an intrusion signal.

For the radiation source signal, taking FSK as an example. It is mentioned above that it uses the frequency change between different pulses to realize signal modulation, that is:
\begin{equation}
s\left (t\right )=Ae^{j\left (2\pi f\left (t\right )+\theta _{0}\right )}+n\left (t\right ), \left (0<t\le T\right )
\label{eq4}
\end{equation}
where $A$ represents the signal amplitude, $f\left (t\right )$ represents the frequency modulation function, $\theta _{0}$ represents the initial phase, and $T$, pulse width.

\begin{figure}[h]
\centering
\includegraphics[scale=0.6]{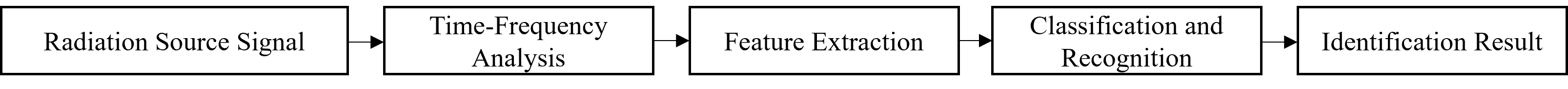}
\caption{Radiation source signal identification model}
\label{fig3}
\end{figure}

Radiation source signal identification can be conducted based on the aforementioned theory. The objective is to receive, process, and identify the radiation source signal, achieving optimal identification results while minimizing errors. A radiation source signal identification model is shown in Figure \ref{fig3}. When employing time-frequency analysis to extract signal features, a higher parameter dimension leads to increased complexity in representing information. Hence, it is crucial to extract parameters with robust explanatory properties in order to effectively identify signal features, which can enhances computational efficiency and improves recognition accuracy during the classification process.

\section{Simulation and Results}
On the basis of the aforementioned theory and technology, the use of radar antenna as a transmitter to spread radio signals carrying computer virus information is demonstrated in this chapter. Typically, a radar detects target by emitting electromagnetic waves (radio waves) and receiving the reflected waves from the target. The emitted signal is modulated and adjusted before being transmitted into space. During transmission, the electromagnetic waves follow the laws of propagation, which are a form of energy propagation resulting from the interaction of electric and magnetic fields, and they can be transmitted wirelessly as radio waves. In radar system, the emitted signal is appropriately modulated in terms of frequency and waveform. Then the modulated signal is transmitted into space through an antenna and propagate as electromagnetic waves. When the electromagnetic waves reach the target or receiver, the receiving system can detect and interpret the electromagnetic signal, extracting the relevant information it carries. Therefore, this process establishes the theoretical foundation for transmitting radar signals through wireless channels in the form of electromagnetic waves.

\begin{figure}[h]
\centering
\includegraphics[scale=0.54]{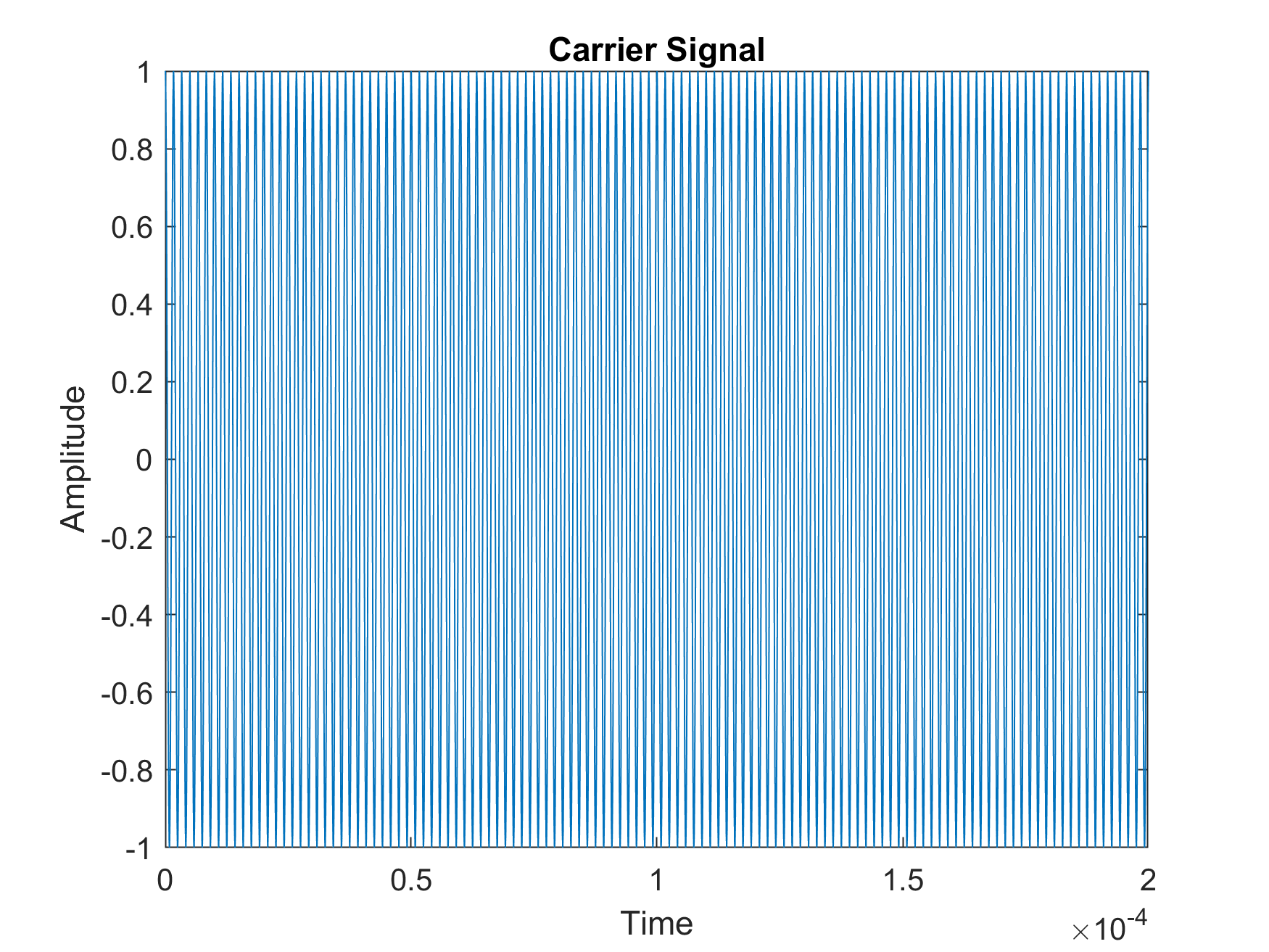}
\hspace{1mm}
\includegraphics[scale=0.54]{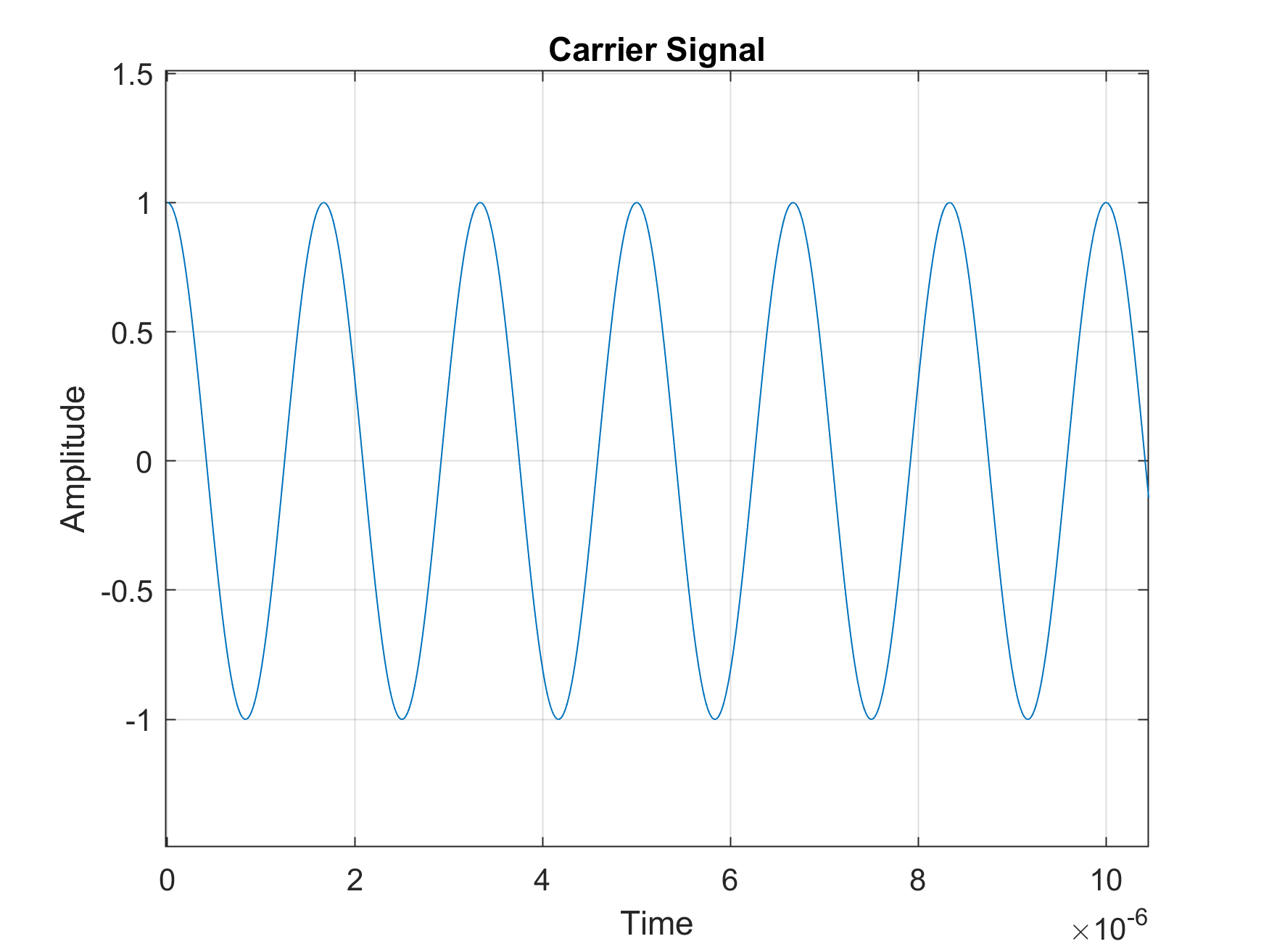}
\caption{Carrier signal. The right subfigure shows the signal waveform of the enlarged part of the signal in the left subfigure.}
\label{fig4}
\end{figure}

For computer virus codes, which are typically expressed in hexadecimal form, it allows for effective utilization of storage space. Moreover, the hexadecimal representation is easier to read and comprehend compared to binary data. By utilizing characters $0-9$ and $A-F$ to represent values $0-15$, it facilitates recognizing the structure and patterns of the code. The utilization of hexadecimal to represent computer virus code serves the purpose of efficiently storing and processing binary data, while also enhancing readability and compatibility.

This paper is interested in the signal processing part. For this, the modulation of a computer virus converted into a binary data stream into a radar carrier signal is simulated. The computer virus is simulated by generating a random binary data stream, and the emission signal is defined by radar parameters. In this case, a radar with a center frequency of 120 GHz is used as the transmitter. In addition, the signal is modulated using FSK to simulate the generation and modulation process of a computer virus signal.

\begin{figure}[h]
\centering
\includegraphics[scale=0.54]{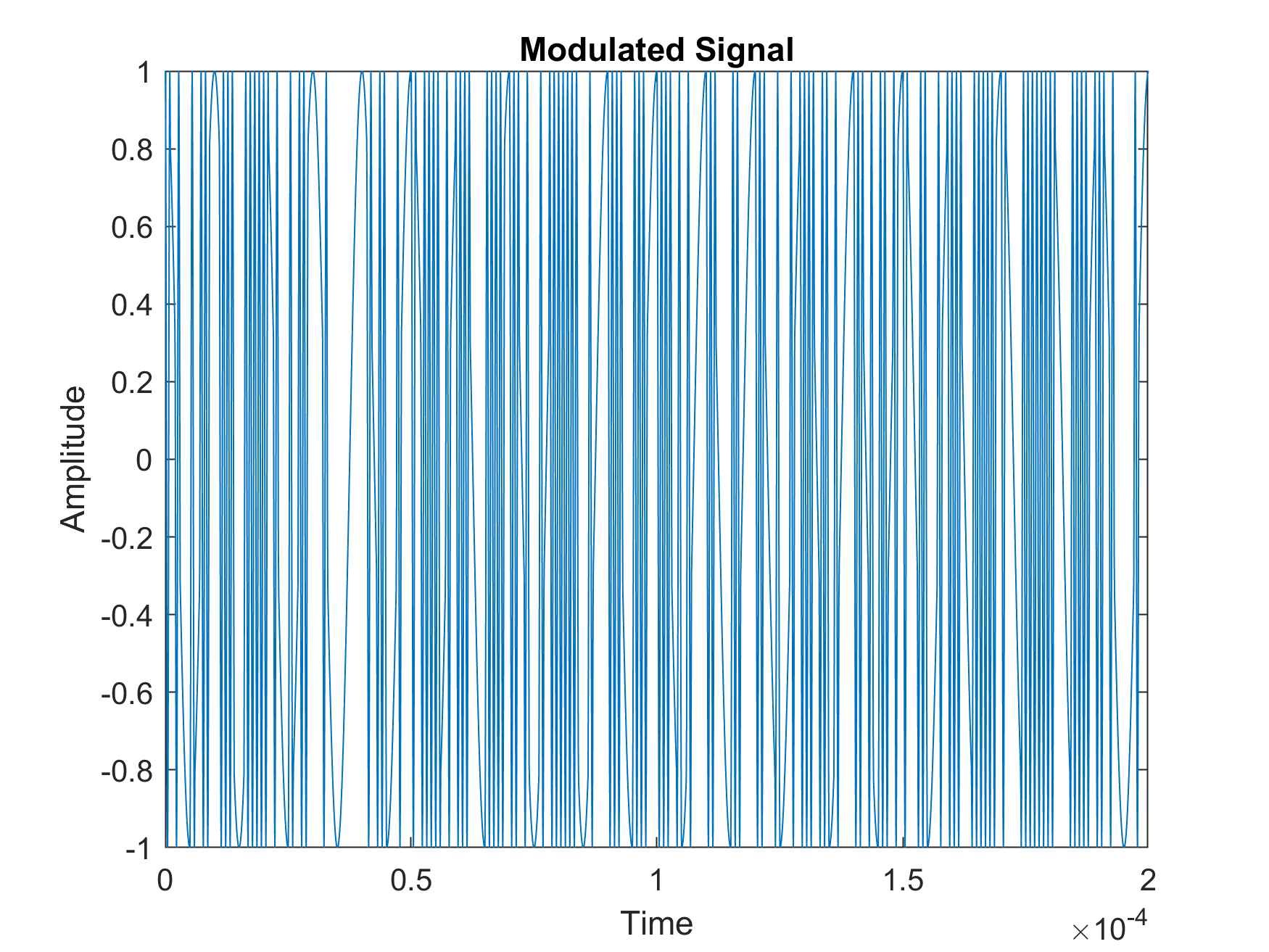}
\hspace{1mm}
\includegraphics[scale=0.54]{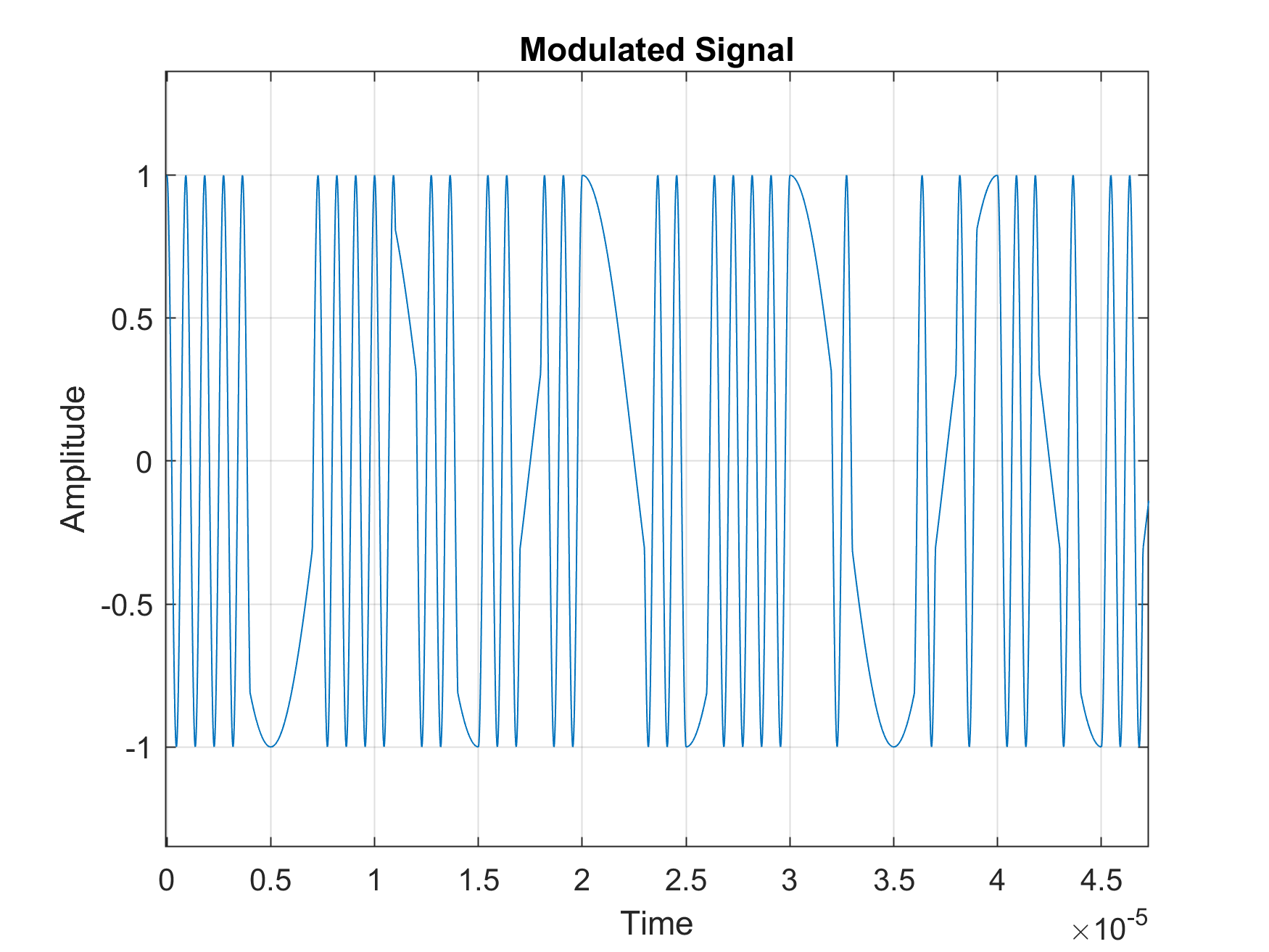}\\
\includegraphics[scale=0.54]{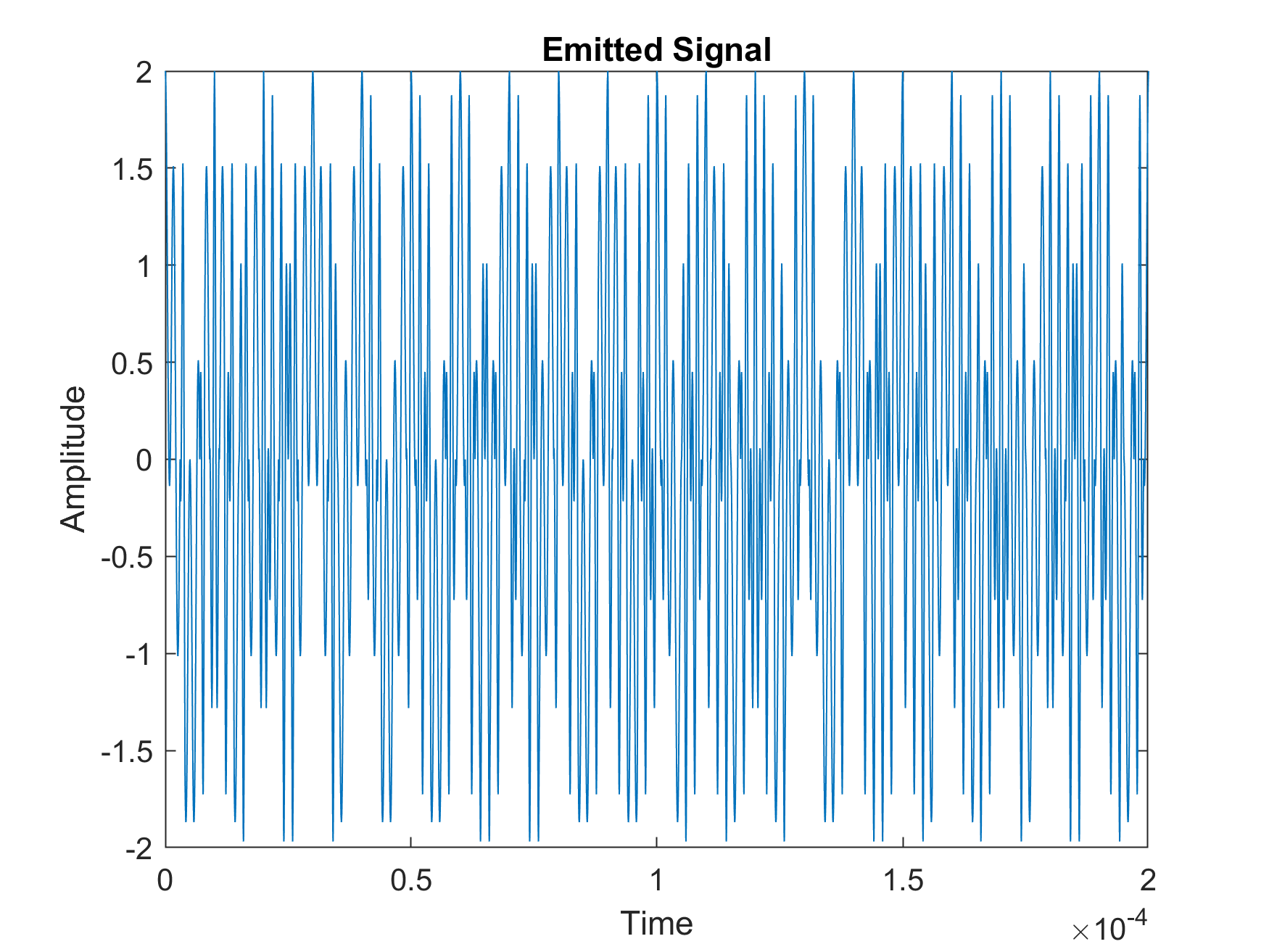}
\hspace{1mm}
\includegraphics[scale=0.54]{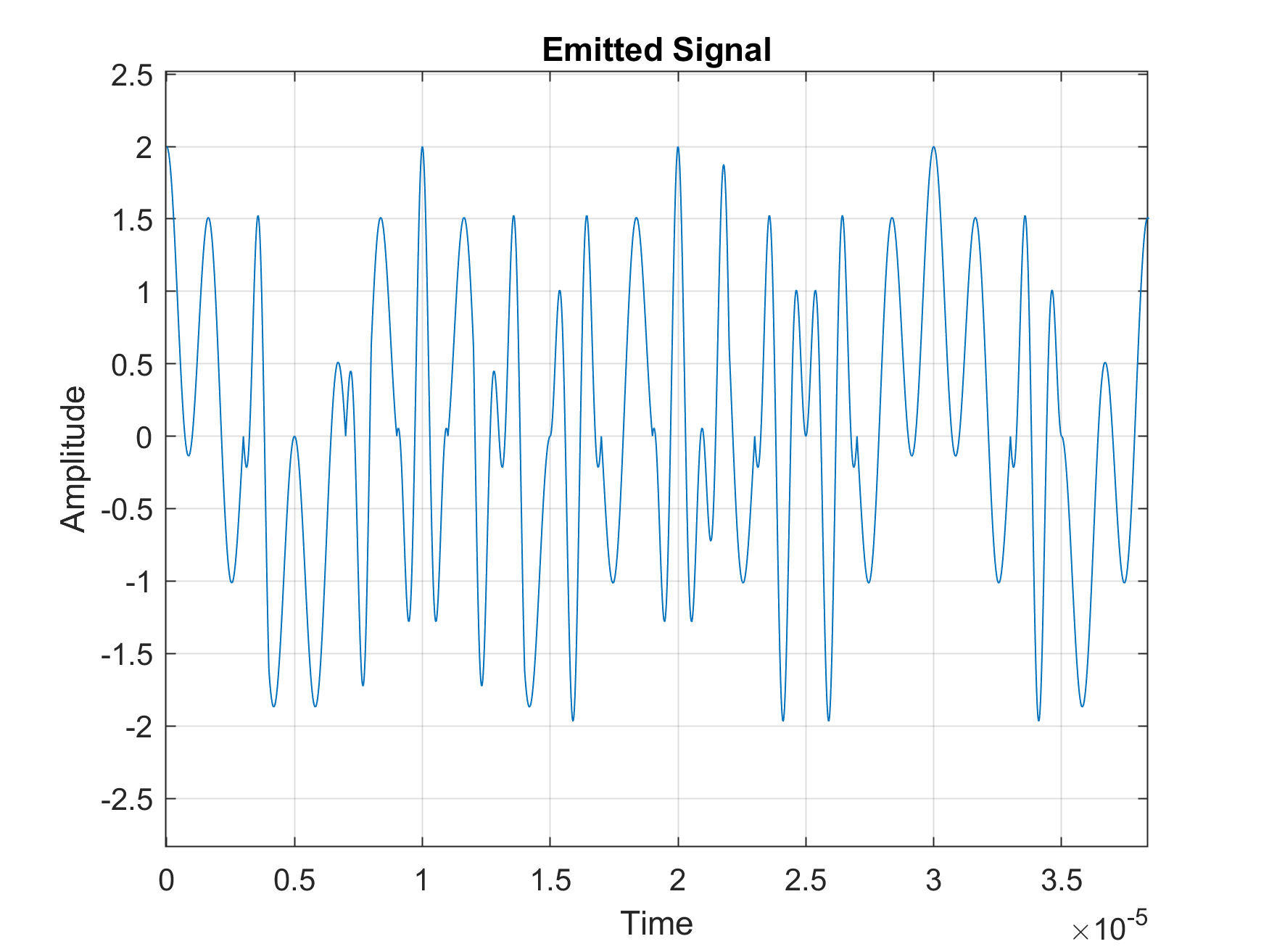}
\caption{Modulated signal and the signal to be emitted. The right subfigures show the signal waveform of the enlarged part of the signals in the left subfigures.}
\label{fig5}
\end{figure}

Figure \ref{fig4} shows a carrier signal with a binary sequence. A cosine carrier signal is generated according to the radar center frequency. Subsequently, the binary data is modulated using FSK by superimposing cosine signals of different frequencies. The frequency of the modulating signal is determined by the value of the data bits. In this simulation, for each binary data bit, if its value is $0$, set the frequency of the carrier signal to $f_{c}-bitRate/2$; on the contrary, set the frequency to $f_{c}+bitRate/2$, that is, a low-frequency cosine signal represents a binary data bit of $0$, while a high-frequency cosine signal represents a binary data bit of $1$. Finally, the modulation signal is added to the carrier signal to generate the signal to be emitted, as shown in Figure \ref{fig5}.

The carrier signal is typically a high-frequency signal, with a frequency much higher than that of the modulating signal. By combining the modulated signal with the carrier signal, the frequency characteristics of the modulating signal can be incorporated into the carrier signal, enabling data transmission and modulation effects. Ideally, the combined transmitted signal can be introduced into the enemy's computer system through radiation coupling, resulting in the release of transmitted data information.

\begin{figure}[h]
\centering
\includegraphics[scale=0.54]{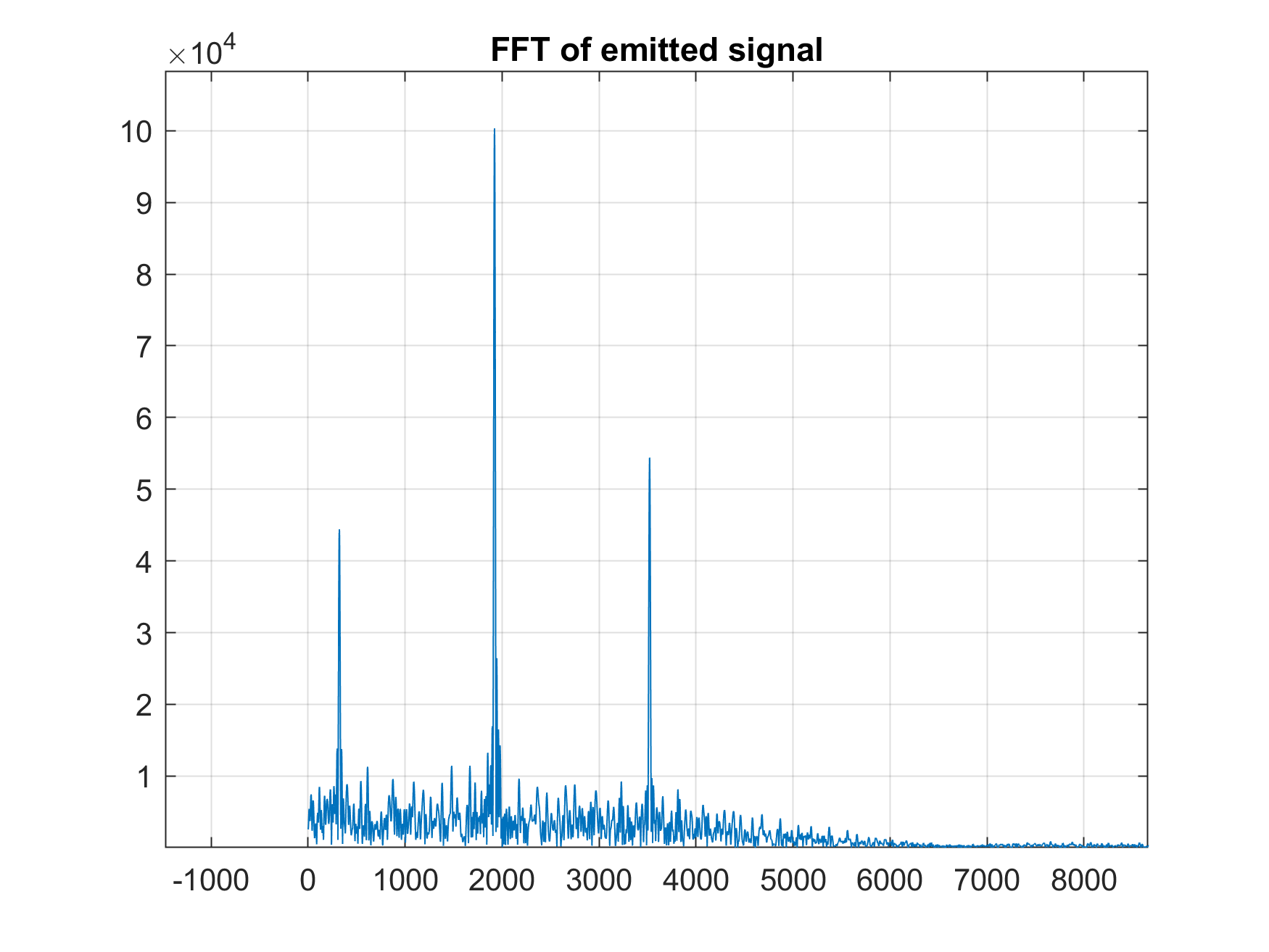}
\hspace{1mm}
\includegraphics[scale=0.54]{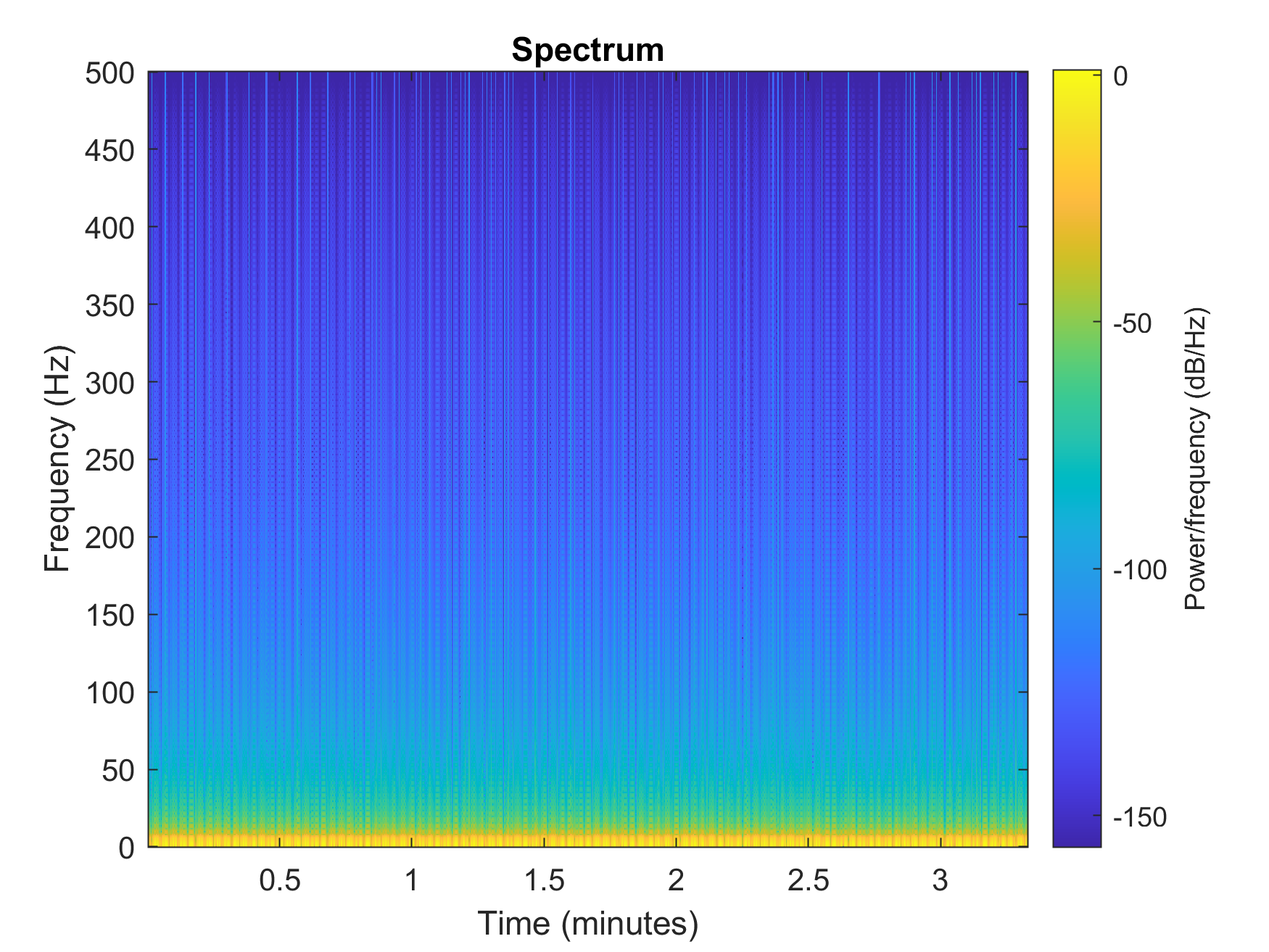}
\caption{Spectrum of emitted signal. The left subfigure shows the FFT of the emitted signal, and the right subfigure, SFT of the emitted signal.}
\label{fig6}
\end{figure}

Figure \ref{fig6} shows the spectrum of the signal intended for transmission, revealing three distinct peaks. During the modulation process of each data bit, different frequency components are employed to generate the modulated signal based on the data bit's value. As a result, the modulated signal contains multiple frequency components. In addition, when the carrier signal is combined with the modulated signal, its frequency aligns with a specific frequency component of the modulated signal. In the SFT spectrum, a vibrant color appears around the frequency of approximately 10 Hz, indicating a high signal strength in this frequency range due to the presence of a powerful signal. Moving from 10 Hz to 500 Hz, the existence of multiple frequency components within this range and a gradual increase in signal energy. Furthermore, at various time points, there are distinct, light-colored straight lines spanning from 0 to 500 Hz, representing instances of higher signal intensity. In this case, the signal power of each frequency component is satisfactory and information attenuation may have little effect on the signal, which lays a solid foundation for the success of the attack mission.

During the actual process, when emitting an electromagnetic wave signal carrying virus information, the radiation information experiences attenuation due to space transmission and "field to line" coupling. To ensure successful injection of the virus into the enemy's computer system, a sufficiently high power is required for the transmitting antenna.

\section{Anti-Radiation Attacks}
With the continuous advancement of attack methods in recent years, relying solely on physical isolation is insufficient to effectively defend against cyber threats. Attacks such as radiation injection have the capability to bypass physical isolation and infiltrate isolated networks to steal information and cause damage. Therefore, in order to effectively identify and respond to these attacks, it is crucial to have a comprehensive understanding of the characteristics of network signals and establish a corresponding virus signal library based on them.

\begin{figure}[h]
\centering
\includegraphics[scale=0.6]{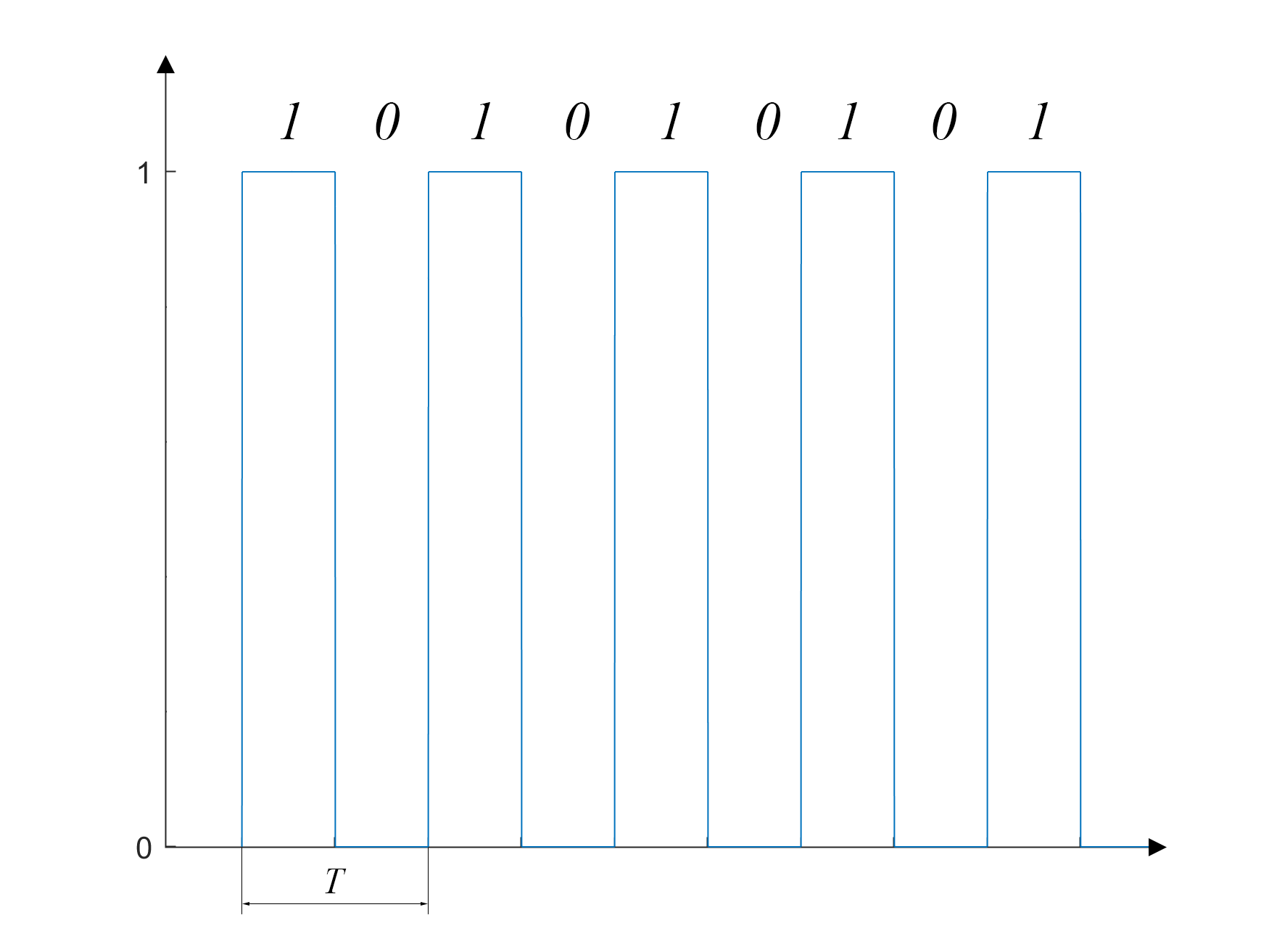}
\caption{Binary signal in time-domain for network transmission}
\label{fig7}
\end{figure}

For a computer to successfully interpret information, it relies on encoding to convert the source information into binary code. This means that the characters or strings processed by the computer are represented by binary numbers, which can be represented by a rectangular waveform, as shown in Figure \ref{fig7}, which presents the time-domain waveform. In actual network communication transmission, the waveform of the network signal resembles a sawtooth wave. This is due to the reciprocal relationship between a sawtooth wave and a rectangular wave. By applying the Fourier transform, the Fourier series of the sawtooth wave can be obtained, which allows to convert the signal from a time domain representation to a frequency domain representation for spectrum analysis.

In network communication, Manchester encoding is commonly used as a coding method to represent binary data, the principle of which is to alter the signal level to represent $0$ and $1$. Each bit time is divided into two equal periods. For a binary $0$, the signal transitions from high to low in the middle of the period, while for a binary $1$, the signal transitions from low to high in the middle of the period. This encoding method offers the advantage of having a level change in each clock cycle, facilitating clock synchronization and minimizing the impact of clock drift on signal demodulation.

In section \ref{sig recogn}, it was mentioned that the spectrum information of the network signal can be used for feature extraction, serving as prior knowledge for the virus signal library. By comparing the spectral characteristics of the injected signals in the network with predefined criteria, their correlation can be verified to identify potential attacks. Leveraging the extensive prior knowledge, the application of AI algorithms like support vector machines (SVM), artificial neural networks (ANN), and deep learning (DL) enables automatic identification and classification of attack signals. This approach can significantly enhance the efficiency of signal recognition, allowing for prompt countermeasures in cases where intrusion behavior is clearly detected.

\section{Conclusion}
This paper examines the feasibility of electromagnetic-based computer virus radiation injection attacks. The characteristics, injection methods, and attack modes of computer virus weaponry is studied. A computer virus propagation model is established to simulate the spread of computer viruses, and the impact of which is evaluated through simulation curves, providing compelling evidence for the viability of radiation attacks. The study further explores virus attack workflow tailored to the characteristics of wireless computer networks and communication protocols. Additionally, the encoding of virus data and signal modulation through signal processing algorithms is analyzed, enabling the generation of virus attack signals, and the process of modulating virus binary code into radar emission signals is simulated. Finally, this paper discusses defense measures based on the radiation injection scheme and proposes a potential approach involving the establishment of a virus signal library and the integration of AI algorithms for detection and analysis, aiming to counteract radiation attacks.

Wireless injection is a promising yet challenging one in the field of cybersecurity. In future research, a comprehensive exploration of network traffic patterns, computer virus data, and machine learning will enable researchers to address current issues. The advancement of intelligent information processing algorithms such as DL will facilitate the development of advanced signal processing for cyber attack and defense.

\bibliographystyle{unsrtnat}
\bibliography{references}

\end{document}